\documentclass[conference,a4]{IEEEtran}
\usepackage{cite}
\usepackage{lipsum}
\usepackage{amsmath,amssymb,amsfonts}
\usepackage{algorithmic}
\usepackage{graphicx}
\usepackage{textcomp}
\usepackage{xcolor}
\usepackage{algorithm}
\usepackage{algorithmic}
\usepackage{multirow}
\usepackage{multicol}
\usepackage{caption}
\usepackage{booktabs}
\usepackage{pifont}

\def\BibTeX{{\rm B\kern-.05em{\sc i\kern-.025em b}\kern-.08em
    T\kern-.1667em\lower.7ex\hbox{E}\kern-.125emX}}
\begin{document}

\title{Adversarial Impacts on Autonomous Decentralized Lightweight Swarms}

\author{
\IEEEauthorblockN{Shaya Wolf}
\IEEEauthorblockA{Computer Science Department\\
University of Wyoming\\
Laramie, WY, USA\\
swolf4@uwyo.edu}
\and

\IEEEauthorblockN{Rafer Cooley}
\IEEEauthorblockA{Computer Science Department\\
University of Wyoming\\
Laramie, WY, USA\\
rcooley2@uwyo.edu}
\and

\IEEEauthorblockN{Mike Borowczak}
\IEEEauthorblockA{Computer Science Department\\
University of Wyoming\\
Laramie, WY, USA\\
mike.borowczak@uwyo.edu}}


\maketitle

\begin{abstract}
The decreased size and cost of Unmanned Aerial Vehicles (UAVs) and Unmanned Ground Vehicles (UGVs) has enabled the use of swarms of unmanned autonomous vehicles to accomplish a variety of tasks. By utilizing  swarming behaviors, it is possible to efficiently accomplish coordinated tasks while minimizing per-drone computational requirements. Some drones rely on decentralized protocols that exhibit emergent behavior across the swarm. While fully decentralized algorithms remove obvious attack vectors their susceptibility to external influence is less understood. This work investigates the influences that can compromise the functionality of an autonomous swarm leading to hazardous situations and cascading vulnerabilities. When a swarm is tasked with missions involving the safety or health of humans, external influences could have serious consequences. The adversarial swarm in this work utilizes an attack vector embedded within the decentralized movement algorithm of a previously defined autonomous swarm designed to create a perimeter sentry swarm. Various simulations confirm the adversarial swarm's ability to capture significant portions (6-23\%) of the perimeter.  

\end{abstract}

\section{Introduction}
Unmanned and autonomous vehicle usage, while increasing in many fields to solve a wide range of problems, can prove problematic without secure decentralized systems. Some industry and business use cases such as search and rescue, firefighting, and environmental clean up have been found to be more efficiently solved using UAV/UGV and autonomous swarms. Smart cities in particular show promise for swarm utilization as many problems must be solved on a city-wide scale and are too time-sensitive and error-sensitive for traditional mechanisms. Swarms allow for scalability and increased reliability while providing resilient problem solving for smart cities. Problems such as crowd control systems, lighting control systems, navigation systems, and military surveillance/protection could all benefit from the use of decentralized swarms of autonomous vehicles. 

A decentralized protocol, SHARKS (Secure, Heterogeneous, Autonomous, and Rotational Knowledge for Swarms), allows for any number of agents to circle around a target (in two dimensions) or ensphere a target (in three dimensions) \cite{ics2-sharks}. SHARKS stabilizes quickly through a decentralized algorithm with two rules. While the initial SHARKS protocol enabled a highly-resilient and efficiently controlled swarm to establish 2D and 3D perimeters, it's susceptibility a coordinated automated  attack was not investigated. In order to evaluate the feasibility and effectiveness of an adversarial swarm, an adversarial  protocol is defined to interfere with the objectives of a SHARKS-enabled swarm. A security metric is defined and the adversarial swarm is assessed based on it ability to influence the SHARKS-enable target swarm.
The remainder of this work is organized as follows: section II summarizes this initial SHARK protocol as well as work related to the intersection of drones, decentralized control and smart cities, section III introduces the adversarial swarm protocol, section IV describes the experimental design utilizing a multi-agent simulator, finally section V and section VI describe the results and related discussion.

\section{Prior Work}

Existing SHARKS-protocol research focused on decentralized navigation of drone swarms in smart city applications without the use of lead nodes in order to mitigate direct cybersecurity threats \cite{ics2-sharks}. The original SHARKS protocol is described here for completeness,

\subsection{SHARKS Protocol}
The SHARKS protocol enables a swarm of agents to evenly disperse and then rotate around the target \cite{ics2-sharks}. During a given computation round, agents move based on two rules related to the location of a target and the directional heading of their nearest neighbor. Each agent moves towards the target (Center Rule) and away from their nearest neighbor (Dispersion Rule). Moving away from their nearest neighbor is done with a certain degree of rotation ($r$) to ensure the agents are capable of rotating around the target.

\subsubsection{Center Rule}
Each agent is programmed to move a specified distance ($\delta$) from a primary target. This enables each agent to stay within a clearly defined distance/range from the target ($\delta \pm \epsilon$). Let $dist$ be the distance between a particular agent and the target, $c$ be the number of units each agent can move per epoch to satisfy the center rule, $loc$ be the agents location in the field, and $empty$ be the case where there are no agents/obstacles at a given location, Algorithm 1 defines the center rule.

\begin{algorithm}[H]
\label{cra}
\caption{Center Rule Algorithm}
\begin{algorithmic}[1]
  \IF {($\delta - dist > \epsilon$)}
    \IF {($loc - c = empty$)}
      \STATE move backwards $c$ units
    \ENDIF
  \ENDIF
  \IF {($\delta - dist < -\epsilon$)}
    \IF{($loc + c = empty$)}
      \STATE move forwards $c$ units
    \ENDIF
  \ENDIF
\end{algorithmic}
\end{algorithm}

\subsubsection{Dispersion Rule}
Each agent also aims to move away from their nearest neighbor, however this is done at an angle $(180+r)^{\circ}$ such that the agents maintain rotation around the target. Again, let $r$ be the rotation applied to each agent, $d$ be the number of units each agent can move per epoch to satisfy this rule, $loc$ be the agents location in the field, and $empty$ be the case where there are no agents/obstacles at a given location, Algorithm 2 defines the dispersion rule.

\begin{algorithm}[H]
\label{dra}
\caption{Dispersion Rule Algorithm}
\begin{algorithmic}[1]
  \STATE Face nearest neighbor
  \STATE Rotate heading/bearing clockwise by 180 + $r$ degrees
  \IF {($loc + d = empty$)}
    \STATE move $d$ units
  \ENDIF
\end{algorithmic}
\end{algorithm}

Since each agent moves $c$ units to satisfy the Center Rule and $d$ units to satisfy the Dispersion Rule each round, the movement of each agent is scaled by the $d$ to $c$ ratio ($d:c$). 
Based on prior experimentation, swarms reach stability quicker by favoring one of these rules over the other instead of weighting each equally.

\subsection{Related Work}

In order to adequately address advanced potential threats in the context of advanced in smart-cities this work looks at recent advances in UAV/UGV and autonomous vehicles technology in smart city applications, navigation technology, decentralized approaches, algorithms without lead nodes, cybersecurity solutions, and research using physical testing scenarios (see summary in Table \ref{RelatedWork}).

\begin{table*}[]\centering
\begin{tabular}{@{}p{.75cm}p{10cm}p{.5cm}p{.5cm}p{.85cm}p{.5cm}p{1cm}p{1cm}@{}}
\toprule
Ref & Focus of Prior Work & Smart City & Nav. Tech. & Decent- ralized  & Lead Nodes &  Cyber security & Physical Testing  \\ \midrule
\cite{Alsamhi_2019'} & Survey, smart city applications for IoT / smart devices & + & + & + & + & + & +   \\
\cite{Andress_2019} & Survey, security issues surrounding drones as well as mitigation techniques & - & - & + & - & + & -   \\
\cite{Alsamhi_2019} & Public safety network between victims and first responders & + & - & - & - & - & -   \\
\cite{Autefage_2014} & Collaboration between two projects for park-cleaning swarm supervision & + & - & - & - & + & -   \\
\cite{Gu_2018} & Real-time modular air pollutant modeling through UAV swarms & + & - & - & - & - & +   \\
\cite{Innocente_2019} & The utilization of decentralized swarm intelligence for autonomous firefighting & + & + & + & - & - & -   \\
\cite{Chaumette_2019} & Control over a drone swarm and location awareness through motion capture  & - & + & - & - & + & +    \\
\cite{Farid_2013} & Review of swarm localization and positioning in closed environments & - & + & - & + & + & - \\
\cite{Demir_2015} & Markov chain based approach for the probabilistic density control of a swarm & - & + & + & - & - & - \\
\cite{Hsieh_2008} & Decentralized controller methods for swarm shape generation & - & + & + & + & - & -  \\
\cite{Kim_2014} & A novel particle swarm optimization framework for decentralized consensus & - & - & + & - & - & -   \\
\cite{Yoshimoto_2018} & A method for swarm navigation using only local data  & - & + & + & + & - &  +  \\
\cite{Ge_2012} & Decentralized clustering and dispersion algorithms for autonomous swarms & + & + & + & - & - & -   \\
\cite{Qui_2015} & Survey on UAV swarms with inspiration from bird flocks & - & + & + & - & - & -   \\
\cite{Hui-Ru_2017} & Optimization data collection technique for drone swarms and sensor networks & + & + & + & + & - & - \\
\cite{Kyoohyung_2018} & A homomorphic encryption scheme to protect drone controllers & - & + & + & - & + & +   \\
\cite{Saska_2016} & Visual-based approach for swarm stabilization with multiple applications & - & - & - & + & - & +   \\ \midrule
 & Frequency Count & 7 & 11 & 11 & 6 & 6 & 6 \\\bottomrule
\end{tabular}
\caption{Related Works Summary}
\label{RelatedWork}
\end{table*}

Smart cities of the not to distance future are likely to contain a plethora of applications for Internet of Things (IoT) devices that support cities in tasks such as communication, transportation, agriculture, monitoring, disaster mitigation, environmental preservation, service delivery, pollution mitigation, energy saving, e-waste reduction, and weather monitoring \cite{Alsamhi_2019'}. With these advances and increased utilization of interconnected mobile drone systems comes the the security issues of airspace interference, surveillance and privacy, DOS/DDoS attacks, scaling issues, information privacy, firmware attacks, and vendor background system attacks as well as proposed mitigation techniques \cite{Andress_2019}. Central to advancing the application and security of our future smart cities are navigation technology, decentralized approaches with and without lead nodes, cybersecurity techniques, and both physical and simulation-based testing.

Prior research has investigated the application of a broad spectrum of technologies to an equally broad spectrum of smart city applications. Prior investigations include the use of public safety network between victims and first responders using space technology, drones, and wearable devices \cite{Alsamhi_2019}, drone-based park cleaning swarm supervision systems \cite{Autefage_2014}, a pollution profiling platform \cite{Gu_2018}, and autonomous and firefighting system \cite{Innocente_2019}. Most of these smart city applications achieved objectives through network that operated without the use of lead nodes. While some testing was done through simulation any focus on security was limited on public safety rather than system-wide cybersecurity and attack resilience.


A significant percentage of a city exists within confined indoor spaces. Prior research by Chaumette et al. investigated the use of indoor drones in adverse locations by utilizing motion capture with synchronous location mapping \cite{Chaumette_2019} while Farid et al. compared various close-environment localization detection and positioning techniques \cite{Farid_2013}.  Regardless of location, a significant issue in autonomous swarms is decentralized coordination and control. Approaches to this problem include probabilistic density control of autonomous swarms \cite{Demir_2015}, controllers for shape generation \cite{Hsieh_2008}, sensor consensus achievement through particle swarm optimization \cite{Kim_2014}, and non-homogeneous robotic swarm navigation \cite{Yoshimoto_2018}.  Probabilistic density control of autonomous swarms has been achieved using Markov chain based approach using a decentralized density computation \cite{Demir_2015}. Decentralized controller methods enable swarms to create two dimensional shapes without requiring inter-agent communication \cite{Hsieh_2008}. Consensus algorithm for distributed sensors through data fusion of neighboring nodes that outperformed classic algorithms by 16.5\% with little to no delay \cite{Kim_2014}. Finally, a leader-follower method for swarm navigation using only local data has been tested with physical drones \cite{Yoshimoto_2018}.

Given the emergent behavior of SHARKS-enabled swarms other bio-inspired algorithms help frame potential applications, possibilities, and limitations. Of particular interest are chaotic behavior systems inspired by ant colonies\cite{Ge_2012} and autonomous flying inspired by bird flocks \cite{Qui_2015}. The ant-inspired algorithm in \cite{Ge_2012} utilizes decentralized clustering and dispersion algorithms for autonomous swarms in two and three dimensions using a chaotic lead node. Finally, drone technology is continuously evolving, generally in order to enhance performance.
Examples of these enhancements include optimization methods for data collection between drone swarms and sensor networks using lead nodes to create a three-level network \cite{Hui-Ru_2017}, homomorphic encryption scheme to protect the arithmetic operations in drone controllers for enhanced security \cite{Kyoohyung_2018}, and self-stabilizing UAV swarm system through visual-based methods rather than global positioning \cite{Saska_2016}. 

Given the current state of the art in smart cities swarms, related technologies, and the lack of focus on security threats, this work provides the following contributions: 
\begin{enumerate}
    \item an explicitly defined security metric for an autonomous swarm protocol (SHARKS)
    \item the definition and implementation of an autonomous adversarial swarm 
    \item quantifiable impact of an adversarial swarm
\end{enumerate}

\section{Adversarial Swarm}

\subsection{Vulnerability}
An autonomous swarm utilizing the SHARKS protocol depends on its neighbors in order to position and rotate itself around the target, but it does not rely on communication between agent, and therefore an agent can not verify the legitimacy of its neighbor. This introduces the possibility for adversarial  agents to infiltrate the swarm and influence its behavior. An adversarial swarm may gain access to the target or interrupt target examination by corralling to swarm and creating a gap in the circling behavior.

\subsection{Security Metric}
Recall that each agent in the main swarm aims to be $\delta$ units away from the center, with agents equally distributed radially to create a sentry rind around the target (target circle). The adversarial  swarm aims to open a gap in the target circle large enough to access the target. A clearly defined security metric is size of the gap created by adversarial swarm. The metric, percent access, is the largest gap created by an adversary normalized by the total circumference of the circle created by all of the agents in the swarm.

\subsection{Adversary Protocol}
The adversarial swarm presented here leverages slight adjustments to the existing SHARKS protocol in order to create an incredibly efficient disruptive swarm. Adversarial swarms that wish to disrupt the SHARKS protocol can enter at a single point and then corral the agents such that access can be granted to the target through the opening created. This adversarial swarm also moves based on two distributed algorithms. Like the original swarm, each agent moves towards the target (Adversarial Center Rule) and away from their nearest neighbor (Adversarial Dispersion Rule) each epoch. This is done with a certain degree of rotation ($r$) to ensure the agents are moving at a reasonable speed around the target. However, adversarial agents ignore any agent in their way, they only disperse from adversarial agents, and they disperse at a slower pace then the original swarm by a factor of .2. 

\subsubsection{Adversarial Center Rule}
The first algorithm is almost identical to the center rule algorithm for our base swarm.  Each agent wants to be $\delta$ units from the target, with a buffer of $\epsilon$ units. Like the original swarm, each adversarial agent stays within a clearly defined distance from the target ($\delta \pm \epsilon$). Let $dist_a$ be the distance between a particular adversarial agent and the target, $c$ be the number of units each agent can move per epoch to satisfy the center rule, $loc_a$ be the adversarial agents location in the field, and $empty$ be the case where there are no agents/obstacles at a given location.

\begin{algorithm}[H]
\caption{Adversarial Center Rule Algorithm}
\begin{algorithmic}[1]
  \IF {($\delta - dist_a > \epsilon$)}
    \IF {($loc_a - c = empty$)}
      \STATE move backwards $c$ units
    \ENDIF
  \ENDIF
  \IF {($\delta - dist_a < -\epsilon$)}
    \IF{($loc_a + c = empty$)}
      \STATE move forwards $c$ units
    \ENDIF
  \ENDIF
\end{algorithmic}
\end{algorithm}

\subsubsection{Adversarial Dispersion Rule}
The dispersion rule is minutely different than the original swarm. Each adversarial agent moves away from the nearest other \textit{adversarial} agent in the field. Like the original swarm, this is done at an angle $(180+r)^{\circ}$ such that the agents maintain rotation around the target. Once again, let $r$ be the rotation applied to each agent, $d$ be the number of units each agent can move per round to satisfy the dispersion rule, and $loc_a$ be the agent's location in the field. Notice that adversaries do not check to see if their destination is empty. 

\begin{algorithm}[H]
\caption{Adversarial Dispersion Rule Algorithm}
\begin{algorithmic}[1]
  \STATE Face the nearest \textit{adversarial} agent
  \STATE Rotate heading/bearing clockwise by 180 + $r$ degrees
  \STATE move ($d$ * .2) units
\end{algorithmic}
\end{algorithm}


\label{experiment}
\section{Experiment}
The following experiment was designed to assess the security impact of various adversarial swarms parameters on a SHARK-enabled swarms.

There were 3,024 unique experiments performed in the Netlogo multi-agent simulator \cite{netlogo_cite}, and each ran three times for a total of 9,072 runs. These experiments consisted of all possible combinations of the following parameters:
\begin{itemize}
\item{\textbf{Population Size: }8, 16, 32, 64} 
\item{\textbf{Number of Adversaries: } 2, 3, 4, 5, 6, 7, 8}
\item{\textbf{Delta ($\delta$): } 8, 12, 16}
\item{\textbf{Epsilon ($\epsilon$): } 4, 6, 8}
\item{\textbf{Algorithms Ratio ($d:c$): }0.5 : 0.5, 0.5 : 1, 1 : 0.5, 1 : 1}
\item{\textbf{Adversary Delay: }No Delay, Stability, Stability + 20}
\end{itemize}

\subsection{Parameters}
\subsubsection{Population Saturation}
The first two parameters considered in this experiment vary the adversarial saturation of the population, or the ratio of adversarial  agents to regular agents. Some of these experiments have populations that are equally saturated with adversaries.

\subsubsection{Stability Region}
The next two parameters affect the shape and area of the stability region. In previous experiments, regions with larger deltas (but the same area) reach stability faster (in experiments without adversaries) since there is a larger circumference for agents to tend to. Notice that there are stability regions with equal areas (but different delta and epsilon values) in these experiments. 

\subsubsection{Agent Movements}
Agent movements are governed by the $d:c$ ratio (the number of units an agent can move to satisfy the Dispersion Rule and the Center Rule) as well as the rotation applied to each agent. Therefore, the $d:c$ ratio configuration is the next parameter. Note that the adversaries and secure agents are tuned the same. In other words, the secure agents and the adversarial  agents have the same $d:c$ ratio and the same rotation value. The same rotation (20 degrees) is applied in all of these experiments.

\subsubsection{Adversary Delay}
The final parameter for these experiments is the amount of time the adversarial  agents must wait before they can launch their attack. Because this attack works by corralling agents, it has varying success if the regular agents have not reached a stable swarming behavior. In each third of the experiments, the adversaries start at the same time as the regular agents (No Delay), the adversaries start when the regular agents reach stability (Stability), and the adversaries start 20 rounds after the regular agents reach stability (Stability + 20).

\subsection{Security Metric Assessment}
Since the aim of the adversarial swarm is to open a gap in the target circle the largest gap opened in the first 10,000 rounds of simulation is divided by the circumference of the target circle in order to obtain the percentage of the circle opened. In the results section, this is referred to as "\% Access" as it is the percentage of the circle that is accessible to the adversarial agents. 

\label{results}
\section{Results}
\subsection{Population Saturation}
\begin{figure}
  \includegraphics[width=\linewidth,height=8cm]{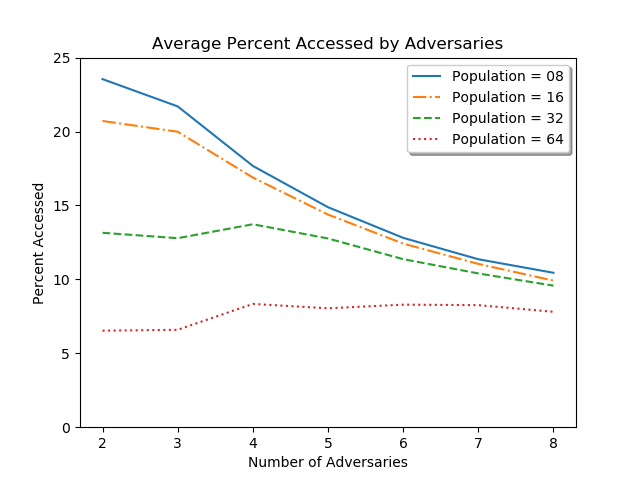}
  \caption{Average percent corralled for each population size and each number of adversarial  agents.}
  \label{PopulationsFigure}
\end{figure}

The results for these experiments can be seen in Figure \ref{PopulationsFigure}. The smaller the regular population, the more corralling the adversaries can do. Also, smaller adversarial  swarms are able to create a bigger opening in the circle because they can move further apart. For instance, with the smallest population (8 agents) and the fewest adversaries (2 agents), the adversarial  agents were able to open up 23.54\% of the circumference, while in comparison, the largest population (64 agents) and the most adversaries (8 agents) were only able to open up 7.81\%. However, for larger populations, the number of adversaries does not have as much effect on the results. In the largest population with the fewest adversaries, the adversarial  agents were still only able to open up  6.53\%. This indicates that this type of attack is more successful against smaller populations and with fewer adversaries. However, when the regular population is larger, more adversarial  agents may be preferable as they perform similarly, but would provide for adversarial resiliency. \\

\subsection{Stability Region}
In previous experiments without adversaries, the size of the stability region effected how quickly the agents were able to reach stability. Larger delta values (ideal distance from the target) had the most effect on the stability speed, as there was a larger circumference for the agents to tend to. This would indicate that swarms running in environments with larger delta values would be more resilient to the adversarial swarm. However, in almost all cases (except for with the largest epsilon value) this is not the case. In fact, the adversaries had similar effectiveness across all delta and epsilon values with all of the experiments averaging to between 11.5\% to 14.5\%. These results are summarized in Table \ref{StabilityTable}. With only three percent difference between these results, further investigation was needed (see the further investigation section). 

\begin{table}[H]
\centering
\begin{tabular}{|c|c|c|c|c|}
\cline{3-5}
\multicolumn{2}{c|}{}                   & \multicolumn{3}{c|}{$\epsilon$}  \\ \cline{3-5} 
\multicolumn{2}{c|}{\multirow{-2}{*}{}} & 4  & 6 & 8 \\ \hline
                                         & 8  & 11.83\%  &  13.02\% & 14.42\%  \\ 
                                         & 12 & 12.51\%  &  13.03\% & 13.25\%  \\ 
\multirow{-3}{*}{$\delta$}               & 16 & 13.30\%  &  13.42\% & 13.96\%  \\ \hline
\end{tabular} 
\caption{Percent Accessed Across Various Stability Regions}
\label{StabilityTable}
\end{table}

\subsection{Agent Movements}
In these experiments, the results were mostly consistent across all of the experiments. However, when the $d:c$ ratio is 1.0:0.5, the adversarial swarm is never able to open up a gap in the circling behavior. Note that a regular agent (in these experiments) can not recognize the difference between an adversarial agent and another regular agent. Since the regular agents are moving twice as far away from their nearest neighbor as they are moving toward the target, they are able to prioritize moving away from an adversary. This indicates that a larger dispersion rule makes the adversary attack less effective. Further, a smaller center rule should result in a less effective attack because the agents stay closer to the ideal distance from the target. These results can be seen in Table \ref{AgentMoveTable}. Notice that for a center rule of 1.0, a larger dispersion rule resulted in a less effective attack and for a dispersion rule of 0.5, a smaller center rule resulted in a less effective attack. 

\begin{table}[H]
\centering
\begin{tabular}{|c|c|c|c|}
\multicolumn{4}{}{} \\
\cline{3-4}
\multicolumn{2}{c|}{\multirow{2}{*}{}} & \multicolumn{2}{c|}{  Dispersion Rule ($d$)  } \\ \cline{3-4} 
\multicolumn{2}{c|}{}              & 0.5  & 1.0     \\ \hline
\multirow{2}{*}{Center Rule ($c$)} & 0.5  & 17.60\% & NULL \\ 
                                   & 1.0  & 18.96\% & 16.22\% \\ \hline
\end{tabular} 
\caption{Percent Accessed Across Agent Movements}
\label{AgentMoveTable}
\end{table}

\subsection{Adversary Delay}

\begin{figure*}
  \includegraphics[width=\textwidth,height=10cm]{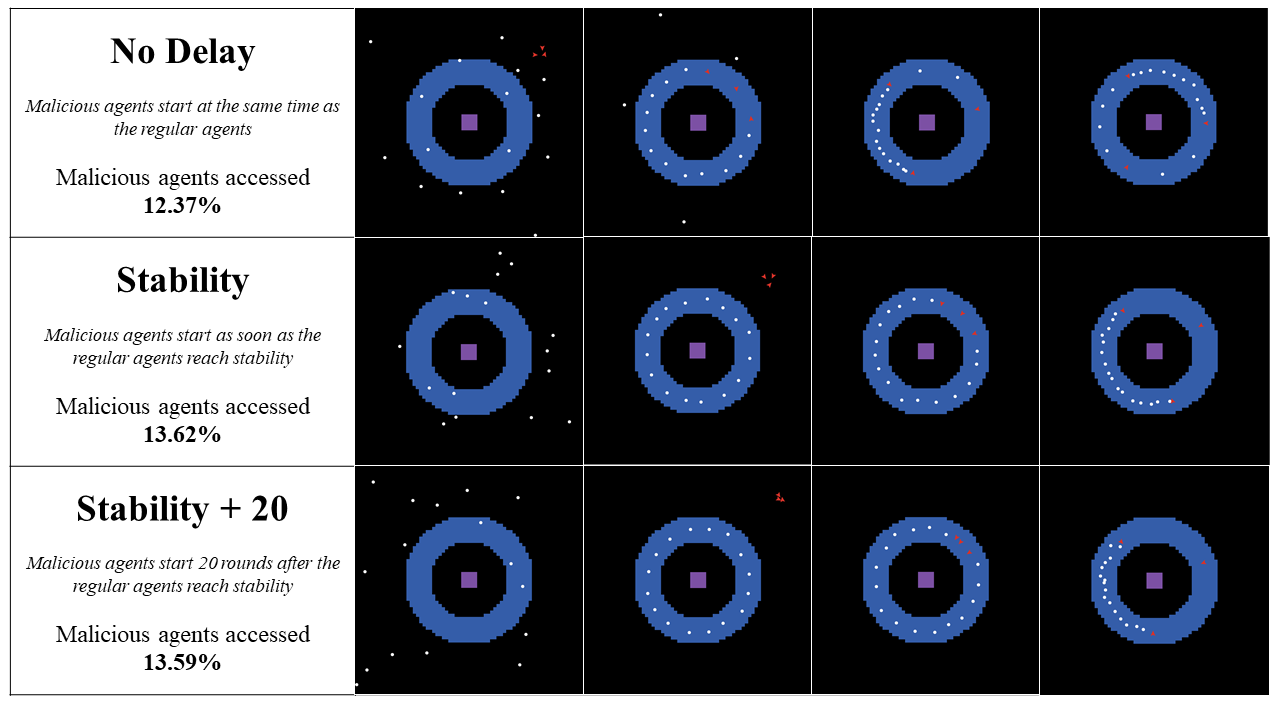}
  \caption{Adversarial swarms movements depending on the amount of delay they invoke. This simulation has 16 regular agents and three adversarial agents. The ideal distance from the target is 16 units $\pm$ 4 units.}
  \label{DelayResultsFigure}
\end{figure*}

Figure \ref{DelayResultsFigure} illustrates the results for adversary delay. When there is no delay, the adversarial  agents start at the same time as the regular agents. The adversaries get dispersed among the regular agents and are unable to create a large opening due to regular agents interrupting their corralling behavior. This results in 12.37\% accessed by the adversaries on average. Recall that the percent accessed means the largest gap between agents relevant to the ideal circumference. So, for example, if an adversarial swarm were to gain 25\% access, there would be a gap spanning 90 degrees of the ideal circle. When the adversaries wait until the regular agents reach stability, they are able to enter the stability region as a unit and start spreading out. The adversaries are able to corral the regular agents until they are equidistant from each other. The largest corralling resulted in 13.62\% accessed. When the adversaries wait 20 rounds after the regular agents reach stability, they have similar results to when they did not add an addition 20 round delay with 13.59\% accessed.

\subsection{Further investigation}
\begin{figure}
  \includegraphics[width=\linewidth]{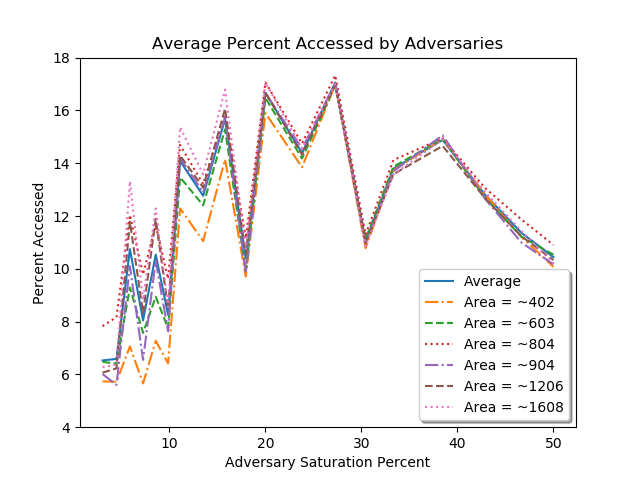}
  \caption{Average percent corralled for different stability region areas.}
  \label{FurtherResults}
\end{figure}

Regardless of this observed behavior, the percent accessed seemed relatively the same across different delays and stability region sizes/shapes. Notice that variations in population saturation created differing results, while the delay results and stability region results looked the same across the different variations (all between 11.5-14.5\%). Data was then analyzed to determine if the population saturation could explain the relatively equal results seen across different types of delays and different stability region configurations. To do this, the percent accessed is plotted against the different population saturation levels for different population sizes. However, this did not show much correlation (seen in Figure \ref{FurtherResults}). 

\begin{figure}
  \includegraphics[width=\linewidth]{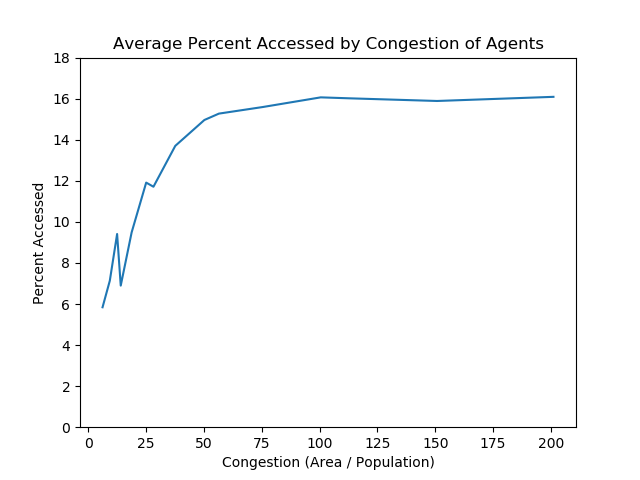}
  \caption{Average percent corralled for different congestion rating (stability region area divided by the population size).}
  \label{CongestionResults}
\end{figure}

Since the different sized areas follow the same pattern and there is a correlation in population size, the final analysis compared the average percent accessed across different levels of congestion. Congestion is defined by the amount of space in the stability region per agent (area of stability region / population size). Notice that smaller congestion ratings mean that the agents are more tightly packed because there is less space per agent. When the agents are more congested, the adversaries attack is less successful. This effect levels off around 75 square units per agent, where the adversarial  agents are able to open around 16\% of the circle. These results are summarized in Figure \ref{CongestionResults}.

\newpage

\section{Discussion}

\subsection{Result Implications}
These results show that under most cases, the adversarial  agents are able to open up sizeable gaps in the swarm circling behavior. Gaps allow for adversaries to access the target as well as prevent comprehensive inspection of the target by the swarm. This could have consequences in smart city applications. For instance, in swarms circling a hazard on a road designed for autonomous vehicles, a gap could allow for an accident to occur. From these results, it is clear that mitigating the risk for this type of attack would mean having more dense swarm populations, which could result in more agent collisions, and smaller delta values, which might not be possible in some environments. Since neither of these solutions are ideal and hinder the flexibility of the SHARKS protocol, more research is needed to provide a way for the SHARKS protocol to neutralize this type of attack. An effective neutralization technique would protect the SHARKS protocol from all corralling attacks.  

\subsection{Limitations}
These results have many limitations including a lack of collision tracking and insufficient data on the size of the opening in units. One of the mitigation techniques involves changing the delta value to make the opening smaller, however this would create a tighter, more congested area for UAVs to move around in. This would likely create collisions between the agents, which is not tracked in these results. These should be tracked in future experiments to show the trade off between this kind of mitigation and the increase in collisions. 

Also, these results are compared across different stability region sizes. To make this consistent across different experiments, the percentage of the circumference that was accessed was recorded. However, this means that equal scores across experiments with different delta values actually create different sized openings. 

Further, these experiments are also run inside virtual simulations. There are additional challenges that would be apparent in real-world experiments. Also, real world applications involve stationary targets as well as dynamic targets that move around an environment. Further testing is needed to test the real-world implications of the SHARKS protocol.

\subsection{Future Work}
Future work on preventing attacks on the SHARKS protocol will look at how to prevent the corralling of agents by adversarial  sources. This could include attempting to more precisely define the possible behaviors of the SHARKS protocol or modifying the original protocol to mitigate the adversarial effects of the adversarial  agents. This research indicates that more congested environments result in neutralization of the attack. Since corralling the regular agents results in more congestion, there is a bit of implicit protection against this type of attack, which is why none of the attacks result in more than 25\% disruption. This could be leveraged to create an addition to the protocol that allows the agents to maneuver around the adversaries when they begin to be corralled. This would amplify the current behavior to neutralize the adversarial effects. 

\section{Conclusion}
This research described a adversarial  attack against the SHARKS protocol in which an adversarial swarm (running a slightly modified version of the SHARKS protocol) is able to disrupt the intended behavior of the SHARKS swarm. This research quantifies the effects of the attack and analyzes which scenarios are more vulnerable to attack. The SHARKS protocol extends on previous research by evaluating potential attack vectors and join the call for inherently secure algorithms by seeking appropriate solutions.

\bibliographystyle{IEEEtran}
\bibliography{00_main.bib}


\end{document}